\documentclass[12pt]{article}
\pdfoutput=1
\usepackage[utf8]{inputenc}
\usepackage[english]{babel}
\usepackage{tabularx}
\usepackage{array}
\usepackage{graphics}
\usepackage[pdftex]{graphicx}
\usepackage{psfrag}
\usepackage{epsfig}
\usepackage{subfig}
\usepackage{amsmath}
\usepackage{mathtools}
\usepackage{amssymb}
\usepackage{bbold}
\usepackage{bbm}
\usepackage{bm}
\usepackage{setspace}
\usepackage{rotating}
\usepackage{colortbl}
\usepackage{longtable}
\usepackage{slashed}
\usepackage{braket}
\usepackage{lineno}
\makeatletter
\usepackage{textcomp}
\usepackage[usenames,dvipsnames]{xcolor}
\usepackage{relsize}
\usepackage{cite}
\usepackage{float}
\usepackage{simpler-wick}
\usepackage{enumerate}
\usepackage{verbatim}
\usepackage{fancyhdr}
\usepackage{multirow}
\usepackage{arydshln}
\usepackage{enumitem}
\usepackage{hyperref}
\hypersetup{colorlinks,citecolor=nicegreen,linkcolor=niceblue}
\hypersetup{colorlinks=true}
\usepackage{pifont}
\usepackage{physics}

\DeclareUnicodeCharacter{2212}{-} 

\allowdisplaybreaks

\setlongtables

\newcommand{\bea}{\begin{eqnarray}}
\newcommand{\eea}{\end{eqnarray}}
\newcommand{\beq}{\begin{equation}}
\newcommand{\eeq}{\end{equation}}
\newcommand{\ec}{\end{center}}
\newcommand{\bc}{\begin{center}}

\newcommand{\pdir}{p\kern -5.2pt\raise 0.2ex\hbox {/}}

\newcommand{\vdir}{v\kern -5.75pt\raise 0.15ex\hbox {/}}
\newcommand{\kdir}{k\kern -5.75pt\raise 0.15ex\hbox {/}}
\newcommand{\epsdir}{\epsilon\kern -5.0pt\raise 0.15ex\hbox {/}}
\newcommand{\bvdir}{\bar{v}\kern -5.75pt\raise 0.15ex\hbox {/}}
\newcommand{\Ddir}{D\kern -7.75pt\raise 0.20ex\hbox {/}}
\newcommand{\Adir}{A\kern -7.75pt\raise 0.20ex\hbox {/}}
\newcommand{\ldir}{l\kern -5.0pt\raise 0.2ex\hbox{/}}
\newcommand{\varepsdir}{\varepsilon\kern -5.5pt\raise 0.15ex\hbox{/}}

\newcommand{\nn}{\nonumber}

\makeatother

\newcommand{\s}[1]{\slashed{#1}}

\definecolor{niceblue}{rgb}{0.15,0.15,0.6}
\definecolor{nicegreen}{rgb}{0.1,0.5,0.1}
\definecolor{Red}{rgb}{1.,0.,0.}

\begin{document}
\unitlength = 1mm

\thispagestyle{empty} 

\begin{center}
\vskip 3.4cm\par
{\par\centering \textbf{\Large\bf Leptonic Meson Decays into Invisible ALP}}
\vskip 1.2cm\par
{\scalebox{.85}{\par\centering \large  
\sc J. Alda Gallo$^{a,b,c}$,  A.~Guerrera$^{c}$, S. Pe\~naranda$^{a,b}$ and S.~Rigolin$^{c}$} \\
{\par\centering \vskip 0.7 cm\par}
{\par\centering \vskip 0.25 cm\par}
{\sl $^a$~Departamento de F{\'\i}sica Te{\'o}rica, Facultad de Ciencias,\\
  Universidad de Zaragoza, Pedro Cerbuna 12,  E-50009 Zaragoza, Spain}
{\par\centering \vskip 0.25 cm\par}
{\sl $^b$Centro de Astropart{\'\i}culas y F{\'\i}sica de Altas Energ{\'\i}as (CAPA), 
  Universidad de Zaragoza, Zaragoza, Spain}
{\par\centering \vskip 0.25 cm\par}
{\sl $^c$~Dipartamento di Fisica e Astronomia ``G.~Galilei", Universit\`a degli Studi di Padova e \\
          Istituto Nazionale Fisica Nucleare, Sezione di Padova, I-35131 Padova, Italy} \\
{\vskip 1.65cm\par}}

\end{center}

\vskip 0.85cm
\begin{abstract}
The theoretical calculation of pseudo--scalar leptonic decay widths into an invisible ALP, $M \to \ell\, \nu_\ell\, a$, is reviewed. 
Assuming generic flavour--conserving ALP couplings to SM fermions and a generic ALP mass, $m_a$, the latest experimental results for 
pseudo--scalar leptonic decays are used to provide updated bounds on the ALP--fermion Lagrangian sector. Constrains on the ALP-quark 
couplings obtained from these channels are not yet competitive with the ones derived from FCNC processes, like $M\to P\,a$ decays. 
These leptonic decays can, however, provide the most stringent model--independent upper bounds on ALP-leptons couplings for $m_a$ 
in the (sub)--GeV range. 
\end{abstract}
\newpage
\setcounter{page}{1}
\setcounter{footnote}{0}
\setcounter{equation}{0}
\noindent

\renewcommand{\thefootnote}{\arabic{footnote}}

\setcounter{footnote}{0}


\newpage

\section{Introduction}
\label{sec:intro}

Light pseudo--scalar particles naturally arise in many extensions Beyond the Standard Model (BSM) of particle physics, as they 
are a common feature of any model endowed with a global $U(1)_{PQ}$ symmetry spontaneously broken at a scale $f_a \gg v$. 
Small breaking terms of the global $U(1)_{PQ}$ symmetry are needed for providing a mass term, $m_a \ll f_a$, to the (pseudo) 
Nambu-Goldstone boson (pNGB). Sharing a common nature with the QCD axion~\cite{Peccei:1977hh,Wilczek:1977pj,Weinberg:1977ma}, 
these class of pNGBs are generically dubbed as Axion-Like Particles (ALPs). The key difference between the QCD axion and a generic 
ALP can be summarized in the fact that ALPs do not need to satisfy the well-known constraint~\cite{Weinberg:1977ma}, $m_a f_a 
\approx m_\pi f_\pi$, that bounds the QCD axion mass and the $U(1)_{PQ}$ symmetry breaking scale via QCD instanton effects. 
Therefore, in a generic ALP framework, one can assume the ALP mass being determined by some unspecified UV physics, and, 
consequently, $m_a$ and $f_a$ can be taken as independent parameters. 

The ALP parameter space has been intensively explored in several terrestrial facilities, covering a wide energy 
range~\cite{Mimasu:2014nea,Jaeckel:2015jla,Bauer:2017ris,Brivio:2017ije,Alonso-Alvarez:2018irt,Baldenegro:2018hng,
Harland-Lang:2019zur,MartinCamalich:2020dfe,DiLuzio:2020oah,Guerrera:2021yss}, as well as by many astrophysical and cosmological
probes~\cite{Cadamuro:2011fd,Millea:2015qra,DiLuzio:2016sbl,Agrawal:2021dbo,Lucente:2021hbp}. The synergy of these experimental 
searches allows to access several orders of magnitude in ALP masses and couplings, cf.~e.g.~Ref.~\cite{Irastorza:2018dyq} 
and references therein. While astrophysics and cosmology impose severe constraints on very light ALPs, the most 
efficient probes of weakly-coupled particles in the MeV-GeV range come from experiments acting on the precision 
frontier~\cite{Essig:2013lka}. Fixed-target facilities such as E949~\cite{Artamonov:2009sz,Artamonov:2008qb,Adler:2008zza}, 
NA62~\cite{CortinaGil:2020fcx,CortinaGil:2021nts} and KOTO~\cite{Ahn:2018mvc} and the proposed SHiP~\cite{Alekhin:2015byh} 
and DUNE~\cite{Kelly:2020dda} experiments can be very efficient to constrain long-lived particles. Furthermore, the rich 
ongoing research program in the $B$-physics experiments at LHCb~\cite{Aaij:2015tna,Aaij:2016qsm} and the $B$-factories 
\cite{Masso:1995tw,Bevan:2014iga,Izaguirre:2016dfi,Dolan:2017osp,Kou:2018nap,CidVidal:2018blh,deNiverville:2018hrc,
Gavela:2019wzg,Merlo:2019anv} offers several possibilities to probe ALP couplings in ALP mass regions not completely 
explored yet. 

The main goal of this letter, is the detailed analysis of pseudo--scalar meson leptonic decays, $M \to \ell\, \nu_\ell a$, 
with an ALP escaping the detector or decaying into an ``invisible'' sector. These decay channels were previously analyzed 
in~\cite{Aditya:2012ay} for a massless ALP and for a universal ALP--fermion coupling. Here, a generic ALP mass and generic, 
yet flavour--conserving, ALP couplings are going to be considered. Moreover, a factor 2 misprint in Eq.~(15) of~\cite{Aditya:2012ay} 
(and equivalently a factor 4 misprint in the hadronic contribution of Eq.~(17) of~\cite{Aditya:2012ay}) is going to be corrected.

%
\section{Leptonic Meson Decays in ALP} 
\label{sec:hadronization}

The most general effective Lagrangian describing ALP interactions with SM fermions, including operators up to dimension 
five and assuming flavor conserving couplings reads: 
\bea
\delta \mathcal{L}^{a,MFV}_{\mathrm{eff}} &=& 
        -\frac{\partial_\mu a}{ 2 f_a} \sum_{i=fer} c_i \,\overline{\psi}_i \gamma^\mu \gamma_5 \,\psi_i 
   \,=\, i \frac{a}{f_a} \sum_{i=fer}  c_i\, m_i \, \overline{\psi}_i \gamma_5 \,\psi_i \,.
\label{eq:ALP_SMMFV}
\eea
The Lagrangian in Eq.~(\ref{eq:ALP_SMMFV}) depends only on nine independent flavor diagonal couplings,
$c_i$, once fermionic vector--current conservation and massless neutrinos are implied. It might be useful,
for simplifying intermediate calculations, and explicitly showing the mass dependence of ALP-fermion couplings,
to write the effective Lagrangian in the ``Yukawa'' basis instead of the ``derivative'' one. The two
versions of the effective Lagrangian in Eq.~(\ref{eq:ALP_SMMFV}) are equivalent up to operators of $O(1/f_a^2)$. 

Using the effective Lagrangian of Eq.~(\ref{eq:ALP_SMMFV}) one can calculate the leptonic decay rates
of pseudo--scalar mesons, $M \to \ell\, \nu_\ell \, a$,
with the ALP sufficiently long-lived to escape the detector without decaying (or decaying into invisible channels). 
In such a case the only possible ALP signature is its missing energy/momentum. In the following, $M_{M}$ and $P_{M}$ 
will denote the mass and 4--momentum of the decaying meson, while leptons and ALP masses and 4--momenta will be 
indicated with $m_\ell$, $m_a$, $p_\ell$, $p_\nu$ and $p_a$ respectively. Neutrinos will be assumed massless. 

\begin{figure}[!t]
\centering
\includegraphics[width=7.05cm, height=6cm, keepaspectratio]{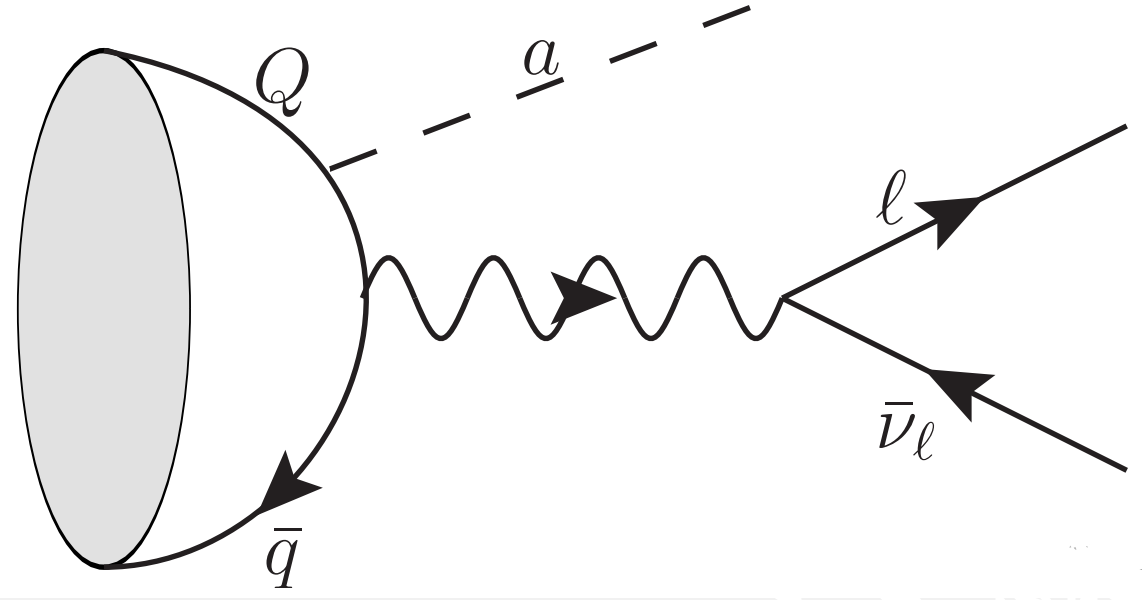}\hspace{1.5 cm}
\includegraphics[width=7.05cm, height=6cm, keepaspectratio]{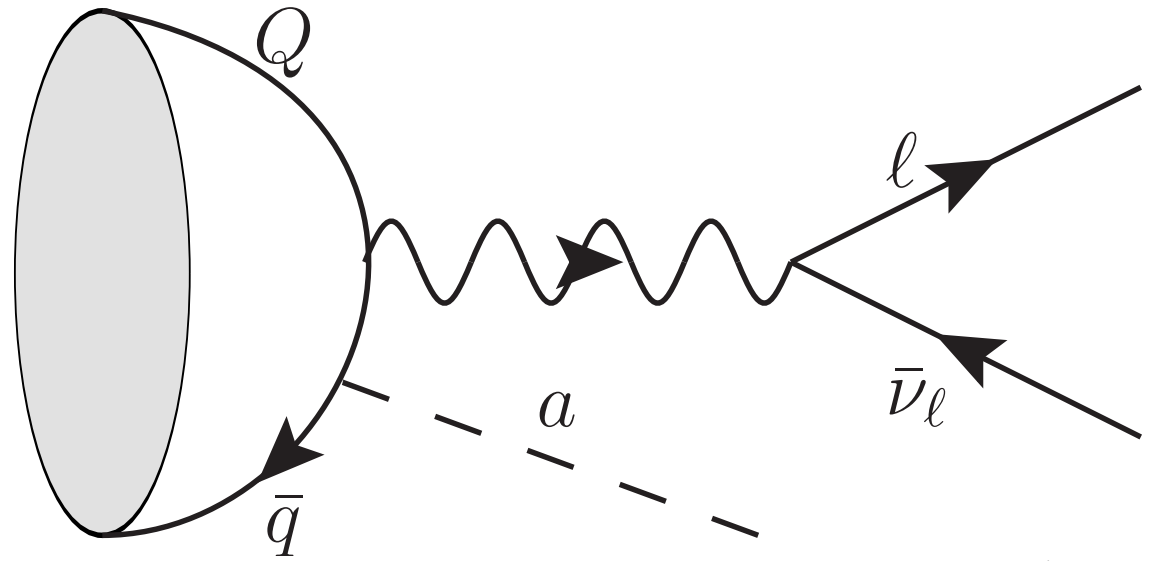}
\caption{Tree level contributions to the $M \to \ell\, \nu_\ell \, a$ amplitude, with the ALP emitted from the $M$ 
meson. The diagram where the ALP is emitted from the charged lepton is straightforward.}
\label{fig:tree_charged}
\end{figure}
Charged pseudo-scalar meson decays proceed through the $s$--channel tree-level diagrams of Fig.~\ref{fig:tree_charged}, 
where only the diagrams where the ALP is emitted from the $M$--meson are shown. The diagram where the ALP is emitted 
from the charged lepton follow straightforwardly, while the one with the ALP emitted from the $W^+$ internal line
automatically vanishes, being the $W^+W^-$--ALP coupling proportional to the fully antisymmetric 4D tensor. In the following, 
the derivation of the decay amplitude for the channel in which the ALP is emitted from the initial quarks or from the 
final charged lepton are discussed separately, as they need two different hadronization treatments.

\subsection{Hadronic ALP Emission}

The two diagrams depicted in Fig.~\ref{fig:tree_charged} represent the contributions to the $M \to \ell\, \nu_\ell \, a$ 
decay in which the parent meson constituent quarks emit the ALP and then annihilate into a virtual $W$ boson, producing 
the final leptons. One refers to this case as hadronic ALP emission. The corresponding amplitude\footnote{For definiteness, 
the leptonic current is written assuming a negative charged meson $M=\bar{q} Q$ state, being $q$ a light up-type quark and 
$Q$ an heavy down-type one.} can be written as:
\bea
\mathcal{M}_h &=&  \bra{0}\bar{q}\,\Gamma^\mu_{h}\,Q\ket{M} \left(\bar{\ell} \,\gamma_\mu P_L \,\nu_\ell \right) \nn \,,
\label{eq:hadronicM}
\eea
with $\Gamma^\mu_{h}$ given by
\bea
\Gamma^\mu_h &=& - \frac{4 G_F}{\sqrt{2}} V_{qQ}
         \left(\frac{c_q\,m_q}{f_a}\,\gamma^\mu P_L\,\frac{\slashed{p}_a-\slashed{p}_q+m_q}{m_a^2 -2 p_a \cdot p_q}\,\gamma_5 
         \,-\, \frac{c_Q\,m_Q}{f_a}\,\gamma_5\,\frac{\slashed{p}_a-\slashed{p}_q-m_Q}{m_a^2- 2 p_a \cdot p_Q}\,\gamma^\mu P_L \right) .
          \label{eq:GammaM} 
\eea
In Eq.~(\ref{eq:GammaM}) $p_q$ and $p_Q$ are the initial quarks momenta, with $c_q$ and $c_Q$ the corresponding ALP-fermion 
couplings.  

The calculation of the $\bra{0} \bar{Q}\, \Gamma^\mu_h \,q\,\ket{M}$ hadronic matrix element in Eq.~(\ref{eq:hadronicM}) 
is complicated by the fact that the meson is a bound state of quarks and one must assume a model to describe the effective 
quark-antiquark momenta distribution. This can be done following the Lepage--Brodsky technique \cite{Lepage:1980fj,
Szczepaniak:1990dt}. In the case of a massless ALP and universal ALP-fermion couplings this amplitude have been firstly 
derived in \cite{Aditya:2012ay}.

Following \cite{Lepage:1980fj,Szczepaniak:1990dt,Aditya:2012ay}, the ground state of a meson $M$ is parameterized 
with the wave--function
\beq
\Psi_M(x)=\frac{1}{4}\phi_M(x)\gamma^5(\s{P}_M + g_M(x) M_M).
\label{eq:wave_parameter}
\eeq
In Eq.~(\ref{eq:wave_parameter}), with $x$ one typically denotes the fraction of the momentum carried by the heaviest 
quark in the meson. The function $\phi_M(x)$ describes the meson's quark momenta distribution, that for heavy and light
mesons reads, respectively:
\bea
\phi_H(x) \propto  \left[\frac{\xi^2}{1-x}+\frac{1}{x}-1\right]^{-2} \qquad , \qquad \phi_L(x) \propto  x(1-x) \,,\label{eq:Wfunction}
\eea
with the normalization fixed such that:
\beq
\int_0^1 dx\,\phi_M(x) = 1.
\label{eq:normalization}
\eeq
The parameter $\xi$ in $\phi_H(x)$ is a small parameter typically of $O(m_q/m_Q)$, being $q$ and $Q$ the light and heavy quark 
in the meson. The mass function $g_M(x)$ is usually taken to be a constant varying from $g_H(x) \approx 1$ and $g_L(x) \ll 1$ 
for a heavy or a light meson. The hadronic matrix element can then be obtained by integrating, over the momentum fraction $x$, 
the trace of the $\Gamma^\mu$ amplitude multiplied by the meson wave--function $\Psi_M(x)$:  
\beq
\bra{0} \bar{q}\,\Gamma^{\mu}\,Q \ket{M} \equiv i f_M \int_0^1 dx\,\mathrm{Tr}\left[\Gamma^{\mu} \Psi_M (x)\right]\,,
\label{eq:mesonic_parametrization}
\eeq
with the meson decay constants $f_M$ defined as: 
\bea
& &\bra{0}\bar{q}\,\gamma^\mu \,\gamma_5 \,Q \ket{M} = i f_M P_M^\mu \,.
\label{eq:Mformfactors}
\eea
In Eqs.~(\ref{eq:wave_parameter}--\ref{eq:mesonic_parametrization}), a slightly different notation with respect to the referred 
literature is used. In particular the functions $\phi_M(x)$ have been normalized to one, in such a way that in 
Eq.~(\ref{eq:mesonic_parametrization}) the mesonic form factor can be explicitly factorized.  

Inserting Eq.~(\ref{eq:GammaM}) and Eq.~(\ref{eq:wave_parameter}) into Eq.~(\ref{eq:mesonic_parametrization}), and defining 
the initial quark momenta as:
\bea
p_q  = (1-x) P_M  \qquad  &,& \qquad  \nn p_Q  = x P_M
\eea
one obtains the following decay amplitudes for the meson ALP--emission process:
\bea
\mathcal{M}_{h} &=& \frac{4\,i\,G_F\,V_{qQ}}{\sqrt{2}} \frac{f_M}{f_a} \frac{M_M^2}{2\,p_a \cdot P_M} \, 
                    \left[ c_Q \frac{m_Q}{M_M} \Phi^{(Q)}_M(m_a^2) - c_q \frac{m_q}{M_M} \Phi^{(q)}_M (m_a^2) \right]
                      \left(\bar{\ell} \, \slashed{p}_a \,P_L \, \nu_\ell \right)
\label{eq:MMALP} 
\eea
where the functions $\Phi^{(q,Q)}_M (m_a^2)$ contain the integrals over the quark momentum fraction
and are defined respectively as:
\bea
\Phi^{(q)}_M (m_a^2) &=& \int^{1-\delta_M}_0 \frac{p_a \cdot P_M}{m_a^2-2\,(1-x)\,p_a\cdot P_M} \, \phi_M(x) \,g_M(x)\, dx \nn \\
\Phi^{(Q)}_M (m_a^2) &=& \int^1_{\delta_M} \frac{p_a \cdot P_M}{m_a^2-2\,x\,p_a\cdot P_M}     \, \phi_M(x) \,g_M(x)\, dx \,.\nn
\eea
The presence of the kinematical cutoff $\delta_M=m_a/(2 M_M)$ prevents the appearance of unphysical bare singularities.

One can check the calculation done in Ref.~\cite{Aditya:2012ay} by taking the $m_a=0$ limit in Eq.~ (\ref{eq:MMALP}) and by setting 
$c_q=c_Q=2$, as demanded by the different normalization of the corresponding ALP-fermion couplings introduced in the effective 
Lagrangians. Notice that 
\bea
\left[\frac{m_b}{M_B} \Phi^{(b)}_B (0)- \frac{m_u}{M_B} \Phi^{(u)}_B (0) \right] = 2\,\sqrt{6}\,\Phi(m_b,M_B) \,
\label{eqPetrov}
\eea
with $\Phi(m_b,M_B)$ the integral defined in Ref.~\cite{Aditya:2012ay}. Doing all these replacements one realizes that Eq.~(15) 
of Ref.~\cite{Aditya:2012ay} is wrong and $1/2$ of the result obtained from Eq.~(\ref{eq:MMALP}).

\subsection{Leptonic ALP Emission}

The leptonic decay amplitude for the lepton ALP--emission process can be easily obtained by using the definition of the meson 
form factors of Eq.~(\ref{eq:Mformfactors}), giving
\bea
\mathcal{M}_\ell &=&  \bra{0}\bar{q}\,\gamma_\mu  P_L\,Q \ket{M} \left(\bar{\ell} \,\Gamma^\mu_\ell \,\nu_\ell \right) \nn \,,
\eea
with 
\bea
\Gamma^\mu_\ell &=& - \frac{4 G_F}{\sqrt{2}} V_{qQ}
\left(\frac{c_\ell\,m_\ell}{f_a}\,\gamma_5\,\frac{\slashed{p}_a+\slashed{p}_\ell+m_\ell}{m_a^2 + 2 p_a \cdot p_\ell}\,\gamma^\mu P_L \right) .
\label{eq:GammaL}
\eea
In Eq.~(\ref{eq:GammaL}) $p_\ell$ dubs the momentum of the final charged lepton. By making use of all the Dirac matrices relations 
one obtains:
\bea
\mathcal{M}_\ell &=& - \frac{4\,i\, G_F}{\sqrt{2}} V_{qQ} \,\frac{f_M}{f_a}  \left[  c_\ell\,m_\ell\, \left(\bar{\ell} \, P_L \,\nu_\ell \right) - 
\frac{ c_\ell \, m^2_\ell}{m_a^2\,+\, 2\,p_a\cdot p_\ell} \left(\bar{\ell} \, \slashed{p}_a\, P_L \,\nu_\ell \right) \right] \,.
\label{eq:MLALP}
\eea
From Eq.~(\ref{eq:MLALP}), by setting $m_a=0$ and $c_\ell=2$ one recovers correctly the result in Eq.~(7) of Ref.~\cite{Aditya:2012ay}.

\subsection{Differential Decay Rate}

For the 3-body decay at hand, and assuming a massless neutrino, one can define the following Mandelstam variables:
\bea
s &=& (P_M - p_\ell)^2 = (p_\nu+p_a)^2 = M_M^2 + m_\ell^2- 2 M_M \omega_\ell \\
t &=& (P_M - p_\nu)^2 = (p_\ell+p_a)^2 = M_M^2 - 2 M_M \omega_\nu \\
u &=& (P_M - p_a)^2 = (p_\ell+p_\nu)^2 = M_M^2 + m_a^2- 2 M_M \omega_a 
\eea
with the energy conservation providing the identity:
\bea
s+t+u = M_M^2+m_\ell^2+m_a^2 \,.\nn
\eea
The differential 3-body decay rate of any scalar particle in its rest frame can be simply written as function of two independent final 
energies $\omega_i$, or equivalently of the two independent Mandelstam variables, as 
\bea
\left(d \Gamma_M \right)_{RF} = \frac{1}{(2 \pi)^3} \frac{1}{8 M_M} |\overline{\mathcal{M}_M} |^2 \, d\omega_e \,d\omega_a = 
                       \frac{1}{(2 \pi)^3} \frac{1}{32 M_M^3} |\overline{\mathcal{M}_M} |^2 \,ds \,du 
\label{eq:diffdecay}
\eea
with $\mathcal{M}_M = \mathcal{M}_\ell + \mathcal{M}_h$. The Feynman amplitude squared reads:
\bea
\hspace{-0.5cm}
|\overline{\mathcal{M}_\ell} |^2 &=& C_M \, c_\ell^2\, \frac{m_\ell^2}{M_M^2} \, 
       \left\{\frac{p_\ell \cdot p_\nu}{M_M^2} + \frac{m_\ell^2}{M_M^2}  \left( \frac{ p_a \cdot p_\nu}{m_a^2 + 2\, p_a \cdot p_\ell}  +  
        m_a^2\,\frac{ p_\ell \cdot (p_a+p_\nu)}{(m_a^2 + 2\, p_a \cdot p_\ell)^2} \right)\right\}  \label{msqlep}  \\
\hspace{-0.5cm}
|\overline{\mathcal{M}_h} |^2 &=& C_M \left[ c_Q \frac{m_Q}{M_M} \Phi^{(Q)}_M(m_a^2) - c_q \frac{m_q}{M_M} \Phi^{(q)}_M (m_a^2) \right]^2 
       \frac{2 (p_a \cdot p_\ell) (p_a \cdot p_\nu) - m_a^2\, p_\ell \cdot p_\nu}{(p_a\cdot P_M)^2} \label{msqhad} \\
\hspace{-0.5cm}
\overline{\mathcal{M}_h} \overline{\mathcal{M}^*_\ell} & = &   C_M \, c_\ell \, \frac{m_\ell^2}{M_M^2}
       \left[ c_Q \frac{m_Q}{M_M} \Phi^{(Q)}_M(m_a^2) - c_q \frac{m_q}{M_M} \Phi^{(q)}_M (m_a^2) \right] 
       \frac{m_a^2\,(p_a \cdot p_\nu + p_\ell \cdot p_\nu)} {(m_a^2 + 2\, p_a \cdot p_\ell) (p_a \cdot P_M)} \quad \label{mlephad} 
\eea
with the overall constant factor defined as:
\bea
C_M = 4\,G_F^2\,|V_{qQ}|^2 M_M^4 \frac{f_M^2}{f_a^2} \,.\nn
\eea

One can notice from Eq.~(\ref{mlephad}), that the mixed product is proportional both to the ALP and the charged lepton masses and, 
consequently, can be neglected either for a massless ALP or for meson decays to a light charged lepton.

The total decay rate, for a general ALP mass, can be obtained by numerically integrating the differential decay rate of 
Eq.~(\ref{eq:diffdecay}) in the kinematically allowed region. On the other hand, the massless ALP limit can be easily integrated 
analytically. By setting $m_a=0$ one obtains:
\bea
\Gamma_{M\to\ell\nu_\ell a} &=& \frac{G_F^2\, |V_{qQ}|^2 M_M^5}{384\pi^2} \frac{f_M^2}{f_a^2} \Big\{ c_\ell^2 \,
\left(2 \rho^2 + 3\rho^4 + 12\rho^4 \log\rho - 6\rho^6 + \rho^8 \right) \, + \,  \nn \\
& + &  \left[\frac{c_Q\,m_Q}{M_M} \Phi^{(Q)}_M (0)- \frac{c_q\,m_q}{M_M} \Phi^{(q)}_M (0) \right]^2\, 
\left(1 − 6\rho^2 - 12\rho^4 \log \rho + 3\rho^4 + 2\rho^6\right)\Big \} .\quad \,
\label{totdecaymassless}
\eea
For $c_\ell=c_q=c_Q=2$ one recovers an agreement with the leptonic part of the decay rate in Eq.~(17) of 
Ref.~\cite{Aditya:2012ay}, while the hadronic part is wrong and 1/4 of the result in Eq.~(\ref{totdecaymassless}), 
consistently with what obtained from the Feynman amplitude check.


\section{Bounds on ALP-fermion couplings}
\label{sec:pheno}

Pseudo--scalar leptonic decay experiments can be used to constraint flavour--diagonal ALP-fermion couplings of Eq.~(\ref{eq:ALP_SMMFV}) 
via the ALP (invisible) decay rate derived in the previous section. Leptonic $B$-decays have been measured at $B$-factories, latest BELLE 
data for electron, muon and tau channel can be found in \cite{Belle:2006tbq,Belle:2019iji,Belle:2012egh}, respectively. Charmed meson 
decays have been measured at BESS (see \cite{Eisenstein:2008aa,BESIII:2013iro,BESIII:2018hhz} for $D$ and~\cite{BESIII:2016cws,BESIII:2019vhn}
for $D_s$ decays respectively) and at BELLE \cite{Belle:2013isi}. Leptonic Kaon decays have been measured by KLOE and NA62~\cite{NA62:2012lny,
KLOE:2007wlh,ParticleDataGroup:2020ssz}. In Tab.~\ref{tab:constraints} available experimental determinations for the leptonic pseudo--scalar 
decay branching ratios are summarized and the lowest order SM predictions are shown for comparison.

\begin{table}[t!]
\centering
\begin{tabular}{|c|c|c|c|}\hline
Channel & SM Branching Ratio & Experiment & Ref.\\\hline
\hline
$B^\pm \to e^\pm \bar{\nu}_e$	&	$8.37\times10^{-12}$ & $< 9.8\times 10^{-7}$ & \cite{Belle:2006tbq}	\\\hline
$B^\pm \to \mu^\pm \bar{\nu}_\mu$	&	$3.57\times 10^{-7}$ & $(5.3\pm2\pm0.9)\times10^{-7}$ & \cite{Belle:2019iji}	\\\hline
$B^\pm \to \tau^\pm \bar{\nu}_\tau$	& $ 7.95\times 10^{-5}$ & $(7.2 \pm2.7 \pm 1.1)\times 10^{-5}$ & \cite{Belle:2012egh}	\\\hline\hline
$D^\pm \to e^\pm \bar{\nu}_e$	& $9.51\times 10^{-9}$	&	$< 8.8\times 10^{-6}$ & \cite{Eisenstein:2008aa} \\\hline
$D^\pm \to \mu^\pm \bar{\nu}_\mu$	& $4.04\times10^{-4}$	& $(3.71\pm0.19\pm0.06)\times10^{-4}$ & \cite{BESIII:2013iro}	\\\hline
$D^\pm \to \tau^\pm \bar{\nu}_\tau$	& $1.08\times10^{-3}$ & $(1.2\pm0.24\pm0.12)\times 10^{-3}$ & \cite{BESIII:2019vhn} \\\hline\hline
$D_s^\pm \to e^\pm \bar{\nu}_e$	& $1.24\times10^{-7}$	& $<8.3\times 10^{-5}$ & \cite{Belle:2013isi} \\\hline
$D_s^\pm \to \mu^\pm \bar{\nu}_\mu$	& $5.28\times 10^{-3}$	& $(5.49\pm0.17)\times 10^{-3}$ & \cite{BESIII:2018hhz}	\\\hline
$D_s^\pm \to \tau^\pm \bar{\nu}_\tau$	& $5.15\times10^{-2}$	& $(4.83\pm0.65\pm0.26)\times10^{-2}$ & \cite{BESIII:2016cws} \\\hline\hline
$K^\pm \to e^\pm \bar{\nu}_e$	& $1.62\times 10^{-5}$	&	$(1.582\pm0.007)\times 10^{-5}$ & \cite{ParticleDataGroup:2020ssz} \\\hline
$K^\pm \to \mu^\pm \bar{\nu}_\mu$	& $0.629$	& $0.6356\pm0.0011$ & \cite{ParticleDataGroup:2020ssz}	\\\hline
\end{tabular}
\caption{Lowest order SM predictions and experimental constraints on the considered $M\to\ell \nu$ decay branching ratios.}
\label{tab:constraints}
\end{table}

The main assumption underlying the following phenomenological analysis is that the ALP lifetime is sufficiently long to escape the 
detector (i.e. $\tau_a\gtrsim 100$ ps) or alternatively that the ALP is mainly decaying into a, not better specified, invisible sector. 
In both cases, the ALP signature is a missing energy/momentum, just as for neutrinos. In this scenario, the simplest way to constrain 
ALP--fermion couplings is then to saturate the 1--$\sigma$ experimental limits on the corresponding leptonic branching ratio adding 
the leptonic ALP decay to the leptonic SM amplitude. No kinematical constraint (2-body vs 3-body decay) is used in the analysis at this 
stage. 

The derived bounds on the $U(1)_{PQ}$ breaking scale $f_a$ are shown in Tab.~\ref{tabfa}. These values have been obtained by setting 
the relevant ALP-fermion coupling to one, with all the others vanishing. The results are provided for two reference values of the ALP 
mass $m_a = 0$ GeV and $m_a=M_M/2$ GeV, showing the variability range that should be expected for a massive vs (almost) massless ALP. 
As an example, the first row in Tab.~\ref{tabfa} should be read as follows: the ``up--quark'' columns represent the $f_a$ limits obtained 
by setting $c_u=1$ and $c_b=c_e=0$ for the two reference values of $m_a$, the ``down--quark'' columns represent the limits obtained by 
setting $c_b=1$ and $c_u=c_e=0$, and finally the values in the ``lepton'' columns are obtained by setting $c_e=1$ and $c_u=c_b=0$.

For heavy pseudo--scalar mesons, such as $B,$ $D$ and $D_s,$ the formulas described in Sec.~2 are straightforward. These mesons are 
very well described by the heavy wave function $\phi_H(x)$ in Eq.~(\ref{eq:Wfunction}), with $g_M=1$. Constituent quark masses 
should be used for partons, instead of bare masses, i.e. $M_M = \hat{m}_Q + \hat{m}_q$ (being $\hat{m}_Q \approx m_Q$) with $Q$ and 
$q$ the heavy and light quark in the meson, respectively. The Kaon sector is more delicate as Kaons cannot be treated fully consistently 
neither as heavy or as light mesons \cite{Brodsky:1981rp}. Therefore, as the Kaon mass is not too far from $\Lambda_{QCD}$, the 
Brodsky--Lepage method introduces larger hadronic uncertainties compared to the heavy mesons case. Here, conservatively, the heavy meson 
wave--function is used\footnote{Using the heavy meson wave--function $\phi_H(x)$ one obtains a decay amplitude roughly $2/3$ of 
one obtained using the light meson wave--function, $\phi_L (x)$. A detailed analysis of the hadronic uncertainties for $K$ decays 
can be found in~\cite{Guerrera:2021yss}.}, with $g_K=1$ and the partonic masses defined as $\hat{m}_u = m_u + \Lambda$ and 
$\hat{m}_s = m_s +\Lambda$ with $\Lambda = (M_K-m_u-m_s)/2$ a parameter of order $\Lambda_{QCD}$. Different choices for $g_K$, lead 
to different limits on $f_a$ that can obtained by a simple rescaling of the ones shown in the last two rows of Tab. \ref{tabfa}, 
i.e. $f'_a = g_K \, f_a$. Therefore, smaller values for $g_K$ result in less stringent bounds for the $U(1)_{PQ}$ scale. 

\begin{table}[t!]
\centering
\begin{tabular}{|c||c|c||c|c||c|c|}\hline
\multirow{2}*{Channel}&\multicolumn{2}{c||}{$f_a$ [MeV] up-quark}&\multicolumn{2}{c||}{$f_a$ [MeV] down-quark} &
\multicolumn{2}{c|}{$f_a$ [MeV] lepton}\\ 
\cline{2-7}&$m_a = 0$ & $m_a = M_M/2$ & $m_a = 0$ & $m_a = M_M/2$ & $m_a = 0$ & $m_a = M_M/2$ \\\hline\hline
$B^\pm \to e^\pm \bar{\nu}_e$	&	2849	&	79	&	3918	&	1294	&	0.50	&	0.13	\\\hline
$B^\pm \to \mu^\pm \bar{\nu}_\mu$	&	6016	&	167	&	8274	&	2723	&	218	&	59	\\\hline
$B^\pm \to \tau^\pm \bar{\nu}_\tau$	&	380	&	6	&	522	&	65	&	200	&	55	\\\hline\hline
$D^\pm \to e^\pm \bar{\nu}_e$	&	5960	&	2130	&	5688	&	858	&	1.99	&	0.53	\\\hline
$D^\pm \to \mu^\pm \bar{\nu}_\mu$	&	3923	&	1370	&	3744	&	559	&	267	&	70	\\\hline
$D^\pm \to \tau^\pm \bar{\nu}_\tau$	&	7	&		&	7	&		&	6	&		\\\hline\hline
$D_s^\pm \to e^\pm \bar{\nu}_e$	&	7921	&	2939	&	8236	&	1870	&	2.47	&	0.66	\\\hline
$D_s^\pm \to \mu^\pm \bar{\nu}_\mu$	&	5487	&	1995	&	5706	&	1284	&	349	&	92	\\\hline
$D_s^\pm \to \tau^\pm \bar{\nu}_\tau$	&	21	&		&	12	&		&	17	&	\\\hline\hline
$K^\pm \to e^\pm \bar{\nu}_e$	&	249144	&	87087	&	169804	&	10176	&	243	&	65	\\\hline
$K^\pm \to \mu^\pm \bar{\nu}_\mu$	&  1744  &	497 	&	1188	&	47	&	321	&	60	\\\hline
\end{tabular}
\caption{Limits on the $U(1)_{PQ}$ scale $f_a$ derived from leptonic pseudo--scalar meson decays, setting the relevant ALP-fermion coupling
  equal to one, with all the other couplings vanishing.}
\label{tabfa}
\end{table}

One can immediately realize that the $f_a$ bounds shown in Tab.~\ref{tabfa} from up-type and down-type ALP-quark sectors are far from 
being competitive with the ones derived from FCNC processes, like $K\to \pi\, a$ or $B\to K\,a$. For example, from \cite{Gavela:2019wzg}, 
one can infer a limit $f_a \gtrsim 10^{9}$ MeV stemming from the top-enhanced penguin contribution, assuming $c_t =1$. Tree--level 
diagram contributions to FCNC processes can provide constraints on lighter quark sectors~\cite{Guerrera:2021yss}, giving limits on $f_a$  
in the range $f_a\gtrsim 10^{6}-10^{7}$ MeV. From $Y(ns)$ decays on can obtain a constraint of the same order for the bottom sector~\cite{Merlo:2019anv}. The only pseudo--scalar meson leptonic channel that provide almost comparable bounds on the quark sector is 
the $K^\pm \to e^\pm \bar{\nu}_e$ decay, while most the other pseudo--scalar leptonic decays provide limits in the ballpark 
$f_a \gtrsim 10^{3}-10^{4}$ MeV for the light lepton decays and $f_a \gtrsim 10^{1}-10^{2}$ MeV for the $\tau$ ones.

Nonetheless, pseudo--scalar meson leptonic decays can be still very useful, as they provide the best present limits on the ALP--lepton 
sector for an ALP with $m_a$ in the (sub)--GeV range, bounding $f_a \gtrsim 10^{2}-10^{3}$ MeV for most of the available channels. 
Typically, the muon sector gives better limits on $f_a$ as it combines experimental data with relatively smaller errors and a not 
too large lepton mass suppression of the amplitude in Eq.~(\ref{msqlep}). The electron sector suffers from a larger mass suppression and typically provides bounds on $f_a \gtrsim 10^{5}-10^{6}$ MeV, with the only exception of the $K^\pm \to e^\pm \bar{\nu}_e$ channel 
benefiting from its highly precise determination\footnote{Recall, however, that caution should be used when handling $K$ data as a larger 
hadronic uncertainty has to be accounted for, unavoidably.}. Furthermore, in this ALP mass range, the results presented here on the 
electron coupling $c_e$ can be complementary with present and future ALP-DM searches like EDELWEISS~\cite{EDELWEISS:2018tde} and 
LDMX~\cite{Berlin:2018bsc} and reactor searches at CONNIE, CONUS, MINE, and $\nu$-cleus~\cite{Dent:2019ueq}. 

\begin{figure}[t!]
\begin{center}
  \begin{tabular}{cc}\hspace{-0.5cm}
\includegraphics[scale=0.46]{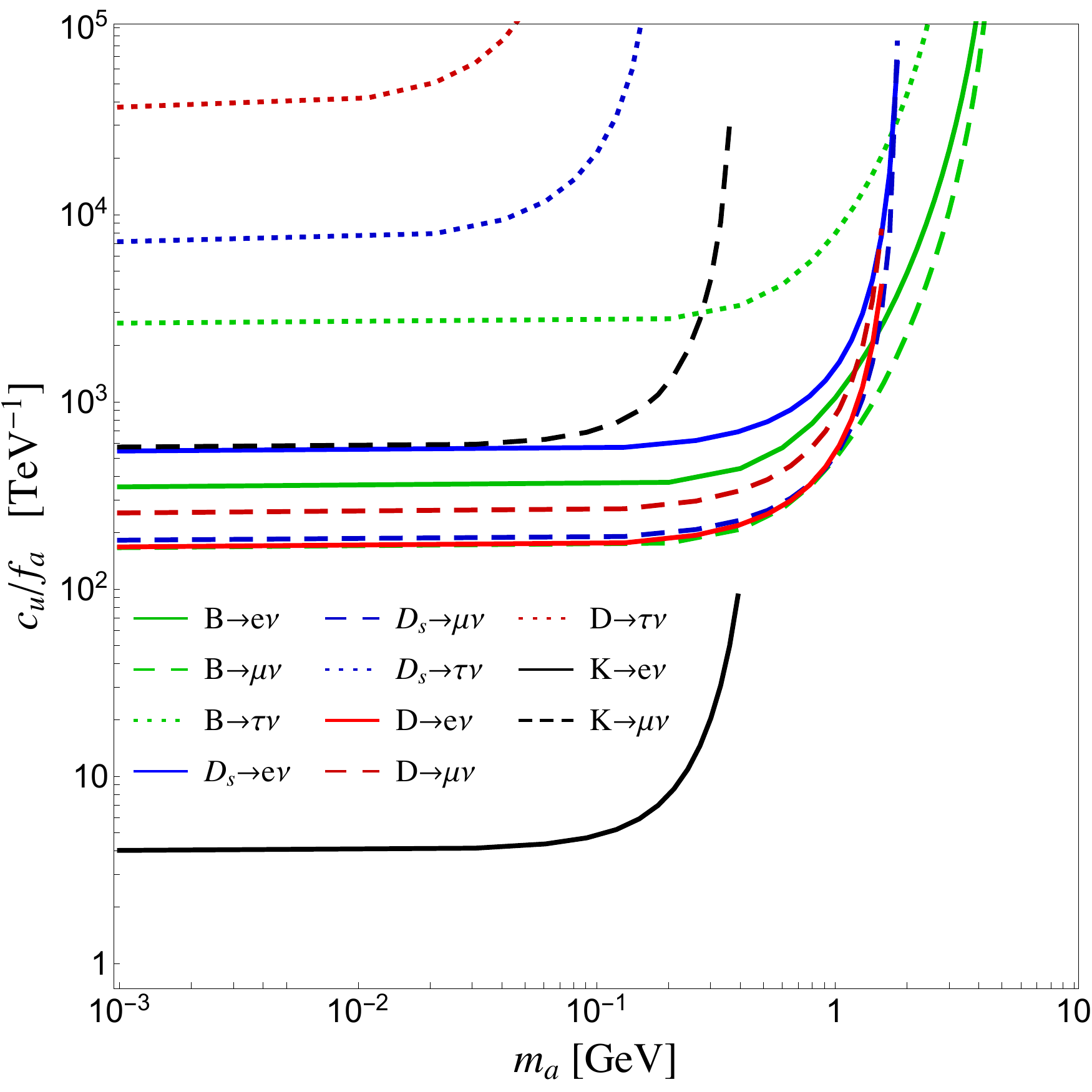} & \includegraphics[scale=0.46]{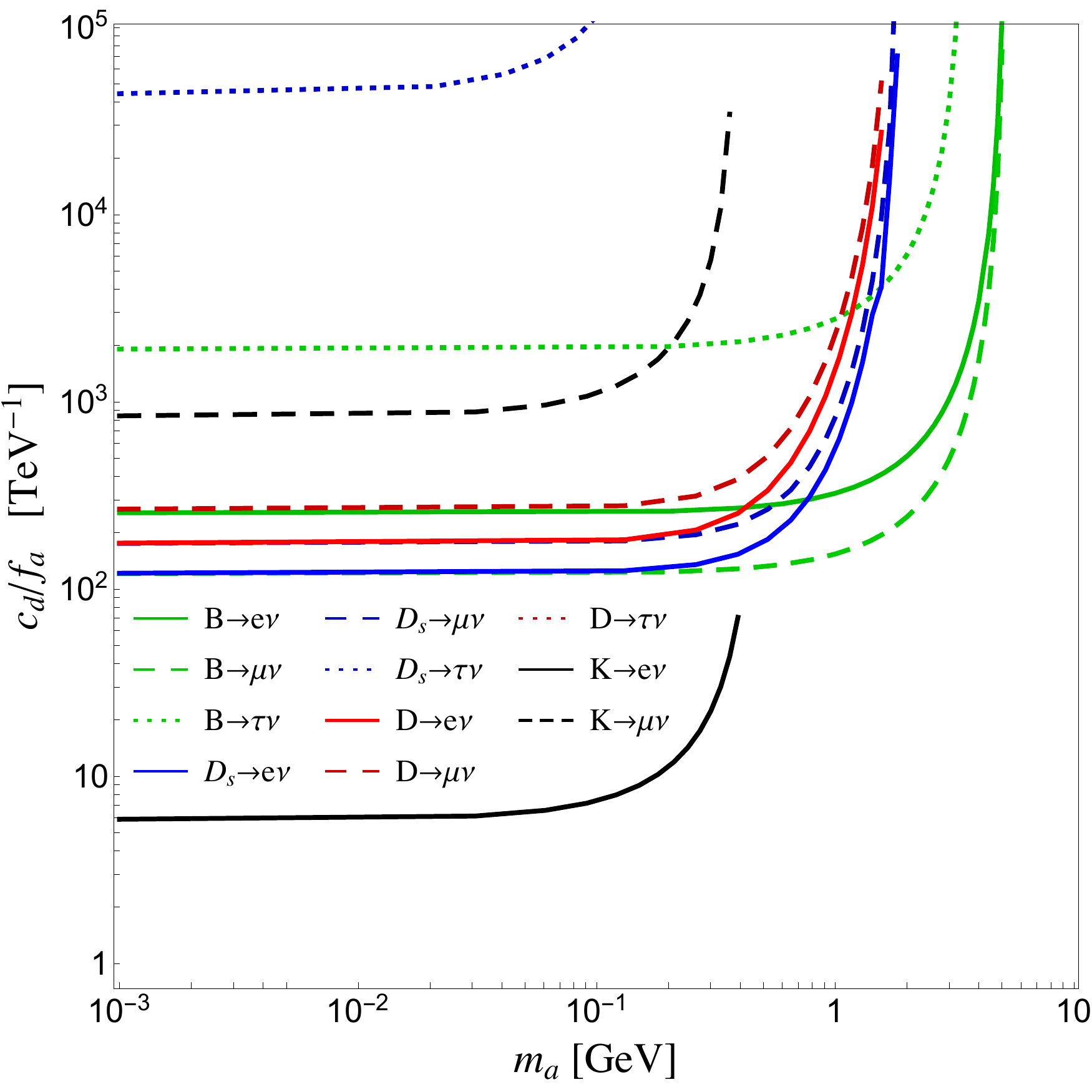} \\ (a)&(b)
   \end{tabular}
\caption{Limits on the coupling (a) $c_u/f_a$ and (b) $c_d/f_a$ derived from the leptonic meson decay indicated in the legend, 
as function of the ALP mass $m_a$.}
\label{fig:quark_plot}
\end{center}
\end{figure}

\begin{figure}[t!]
\centering
\begin{tabular}{cc}
  \hspace{-0.5cm}\includegraphics[scale=0.46]{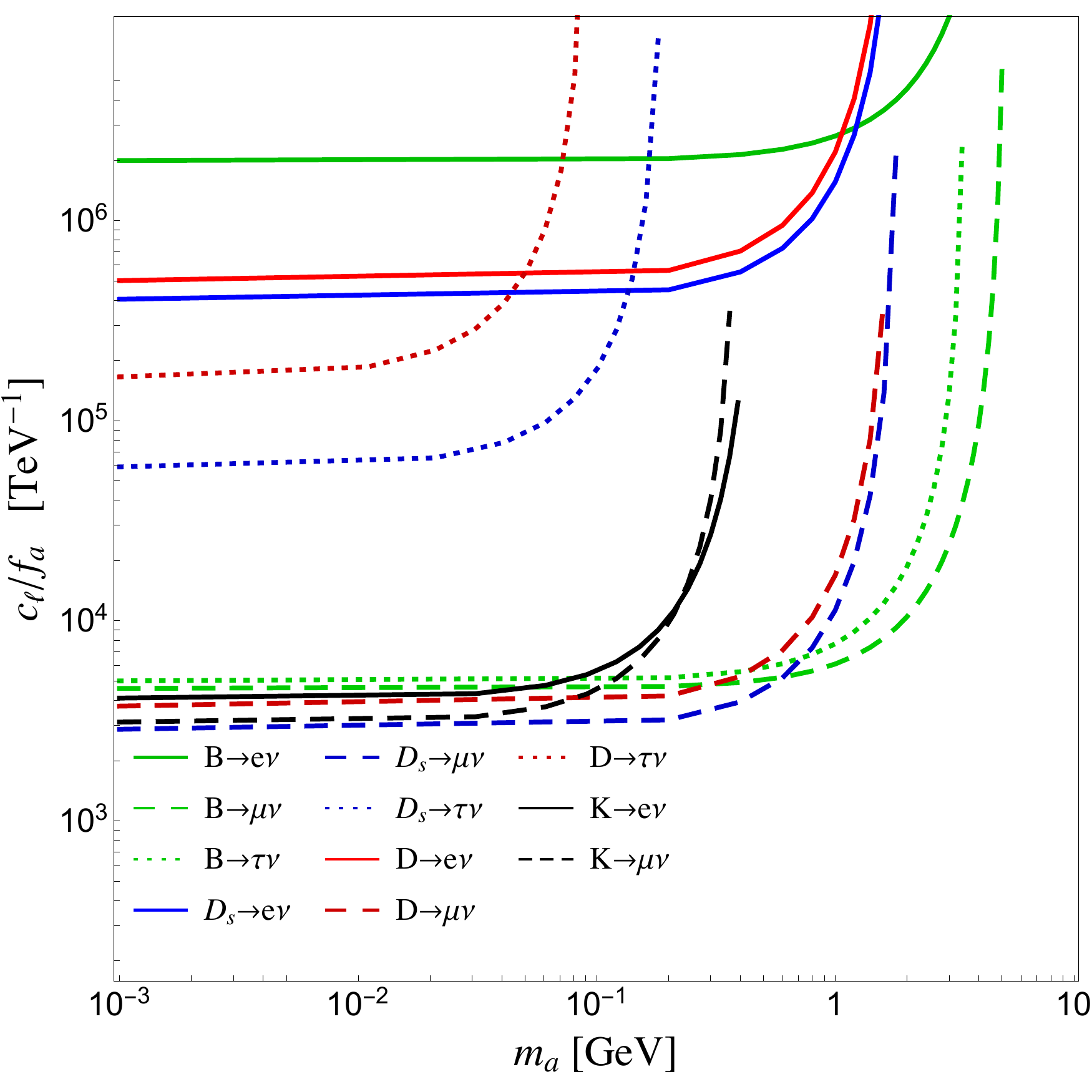} & \includegraphics[scale=0.46]{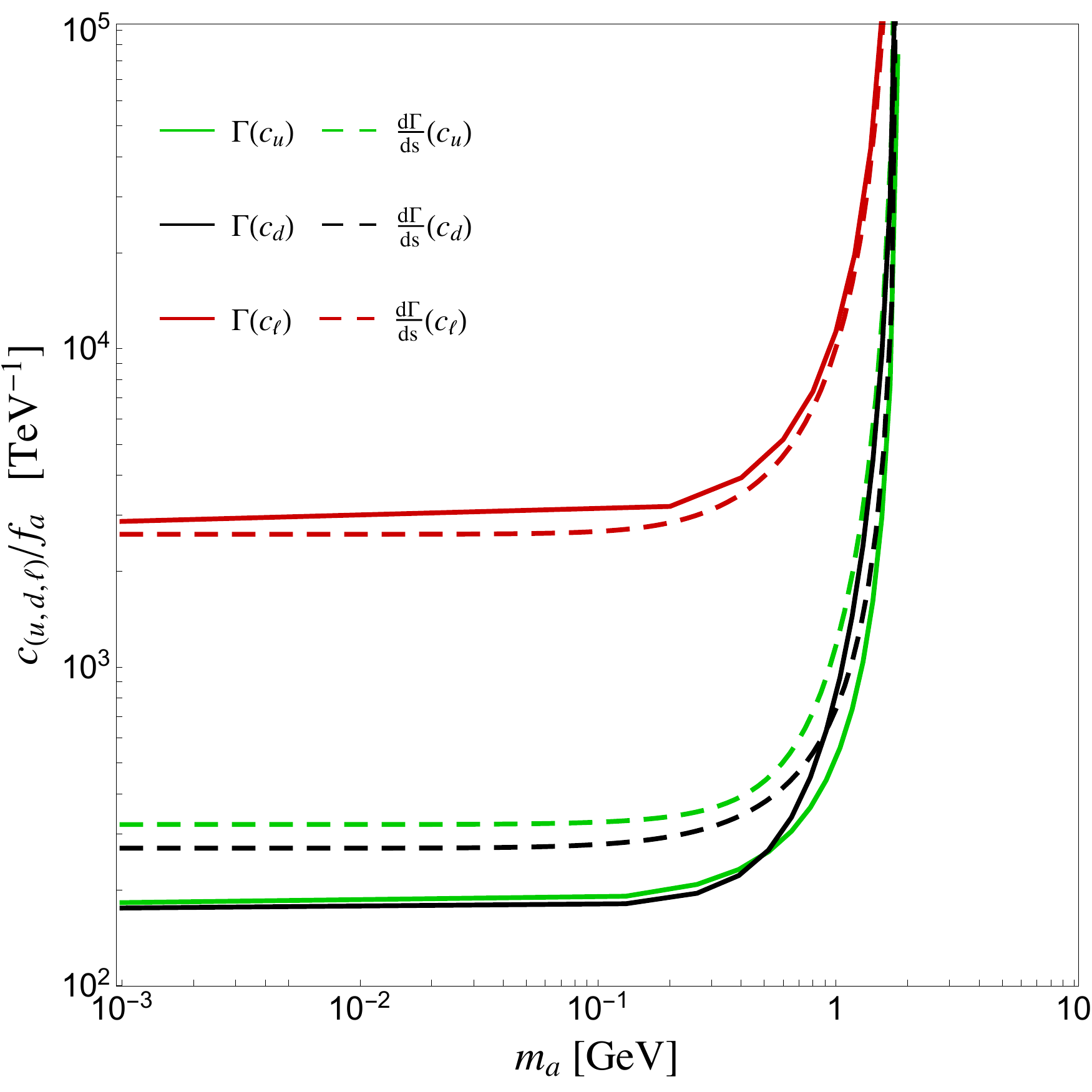} \\ (a)&(b)
\end{tabular}
\caption{Limits on the coupling $c_\ell/f_a$ (a) derived from the leptonic meson decays indicated in the legend, as function of the
  ALP mass $m_a$. Figure (b) shows the limits obtained on all the couplings from the analysis of the $D_s\to \mu\nu_\mu a$
  decay using the experimental BR (full lined) and the missing mass distribution (dashed line). }
\label{fig:lept_plot}
\end{figure}

The same information can be visually obtained from the plots in Fig.~\ref{fig:quark_plot} and Fig.~\ref{fig:lept_plot}, where 
the dependence of the $c_i/f_a$ bounds on the ALP mass is shown for the ALP couplings to up-type and down-type quarks 
(Fig.~\ref{fig:quark_plot} (a) and Fig.~\ref{fig:quark_plot} (b) respectively) and for the ALP couplings to charged leptons 
(Fig.~\ref{fig:lept_plot} (a)). As previously noticed, the $K^\pm \to e^\pm \bar{\nu}_e$ channel is the most promising one, 
putting bounds on $c_{u,s}/f_a \lesssim 5$ TeV$^{-1}$, while most of the other channels are providing limits $c_{u,c,s,b}/f_a 
\lesssim 10^{2}-10^3$ TeV$^{-1}$, still far from the perturbativity region for $f_a=1$ TeV. Concerning the ALP-charged lepton 
coupling notice that the best limits come form $\mu$ decay channels, bounding $c_{\mu}/f_a \lesssim 10^{3}-10^{4}$ TeV$^{-1}$. 
Measures of $c_{\tau}$ are still limited by worse experimental resolution providing bounds $c_{\tau}/f_a \lesssim 10^{5}$ 
TeV$^{-1}$. Sensitivity to the ALP-electron coupling $c_e$ is obviously suppressed by the tiny electron mass giving $c_{e}/f_a 
\lesssim 10^{6}-10^{7}$ TeV$^{-1}$.

The results presented here represent an improvement of at least one order of magnitude compared with limits obtained in Tab. III of~\cite{Aditya:2012ay}. Three main reasons can be advocated: i) first of all, since the publishing of~\cite{Aditya:2012ay}, experimental 
determination of pseudo--scalar leptonic decays has typically improved by roughly a factor ten, leading to more stringent bounds 
on $f_a$, ii) moreover, one has to recall that the leading hadronic contribution in Eq. (17) of~\cite{Aditya:2012ay} underestimates 
by 1/4 the ALP branching ratio, resulting again in lower $f_a$ bounds, iii) finally, assuming a universal ALP-fermion coupling 
results in a parametric cancellation, clearly shown in Eq.~(\ref{eq:MMALP}) and Eq.~(\ref{eqPetrov}) once $c_q=c_Q$ is 
assumed, causing a lost in sensitivity that numerically can be estimated in the 50\%--70\% range\footnote{A detailed and more 
qualitative discussion of this effect can be found in \cite{Guerrera:2021yss}}. 

All the bounds shown up to now have been extracted using only information inferred from the total decay rate. One may 
think that stronger constraints should be derived from the differential decay rate $d\Gamma/d\omega_e$ (or equivalently 
$d\Gamma/ds$) obtained integrating Eq.~\eqref{eq:diffdecay} over the ALP energy $\omega_a$ (or over the Mandelstam 
variable $u$), thus exploiting the different leptonic energy distribution characterizing two--body vs three--body decays. 
The SM two-body decay distribution is peaked around vanishing missing mass $s = m_\nu^2 \approx 0$, and therefore any 
excess of events with $s > 0$ could be an indication of a three-body decay. Unfortunately this analysis cannot be performed 
for most of the decays under considerations as available public results lack of the needed information regarding signal 
and background differential distributions. However, as an example, in Fig.~\ref{fig:lept_plot}(b), the limit on the 
$c_i/f_a$ coefficients obtained from the differential decay rate analysis for the $D_s\to \mu\,\nu_\mu$ decay observed by 
BESIII \cite{BESIII:2018hhz} is shown. BESSIII collaboration provides data on missing mass distribution (i.e. $s$ in our 
notation) only for $s<0.2~\mathrm{GeV}^2$, thus all limits for $m_a > 0.44$ have been obtained assuming a flat background 
distribution up to the kinematical allowed bound. For comparison, in the same plot, also the bounds from the branching 
ratio (solid lines) are reported. 
The analysis reported in Fig.~\ref{fig:lept_plot}(b) should be considered as a theoretical exercise, offering nevertheless 
an order of magnitude comparison between the two approaches, showing that at the moment no clear improvement is obtained 
adding spectral information. Having said that, a more serious effort could be done only having full access to all the 
experimental data of signal and background distributions, and is beyond the scope of this letter.


\section{Conclusions}
\label{sec:conclu}

A detailed analysis of the pseudo--scalar meson leptonic ALP decays, $M \to \ell \, \nu_\ell\, a$ has been presented. 
These decay channels were previously analyzed in Ref.~\cite{Aditya:2012ay} but only for a massless ALP and for a 
universal ALP--fermion coupling. Moreover, a factor 2 misprint in Eq.~(15) of Ref.~\cite{Aditya:2012ay} (and equivalently 
a factor 4 misprint in the hadronic contribution of Eq.~(17) of Ref. \cite{Aditya:2012ay}) has been addressed. 

Bounds on flavor diagonal ALP--fermion couplings are derived from the latest experimental limits on the corresponding 
leptonic decays. The stringent bounds on ALP-quarks couplings can be derived from the $K \to e\, \bar{\nu}_e \,a$ decay, 
with $c_{s,u}/f_a$ around $5$ TeV$^{-1}$, barring large hadronic uncertainties. This bound is, however, still quite 
far from being competitive with the ones derived from the $K \to \pi \,a$ process (see for example~\cite{Guerrera:2021yss} 
for a recent analysis). From heavier pseudo--scalar meson decay channels with a final electron o muon, one can derive 
bounds on ALP-quarks couplings, $c_q/f_a \gtrsim 10^{2}$ TeV$^{-1}$. Typically, less stringent bounds can be obtained 
from the tau channels, mainly due to larger experimental uncertainties. 

Nevertheless, pseudo--scalars leptonic decays can provide the most stringent independent upper bounds on ALP--leptons couplings, 
for and ALP mass,  $m_a$, in the (sub)--GeV range. From $D_s$ and $B$ muon and tau decays one derives limits on $c_{\mu,\tau}/f_a$ 
around $5\times 10^{3}$ TeV$^{-1}$, in all the kinematically allowed $m_a$ range. The most stringent limit on the  ALP--electron 
coupling can be derived from the $K \to e\, \nu_e\, a$ decay, $c_e/f_a \lesssim 4 \times 10^{3}$ TeV$^{-1}$, for $m_a \lesssim 0.3$ GeV. 
For heavier ALP, $D_s$ and $B$ pseudo--scalar meson decays provide much softer bounds with $c_e/f_a \lesssim 10^{6}$ TeV$^{-1}$. 
Present bounds on ALP--electron couplings can be complementary to those obtained from ALP--DM searches \cite{Fortin:2021cog}.

\section{Acknowledgements}

The authors thank Javier Redondo, Maurizio Giannotti and Luca di Luzio for helpful comments and discussions.
A.G. and S.R. acknowledge support from the European Union’s Horizon 2020 research and innovation programme under the Marie 
Sklodowska-Curie grant agreements 690575 (RISE InvisiblesPlus) and 674896 (ITN ELUSIVES). This project has  received 
funding/support from the European Union’s Horizon 2020 research and innovation programme under the Marie Sklodowska-Curie 
grant agreement No 860881-HIDDEN.
The work of J.~A. and S.~P. is partially supported by Spanish grants MINECO/FEDER grant FPA2015-65745-P, PGC2018-095328-B-I00 
(FEDER/Agencia estatal de investigaci{\'o}n) and DGIID-DGA No. 2015-E24/2. J.~A. is also supported by the 
Departamento de Innovaci\'on, Investigaci\'on y Universidad of Arag\'on government, Grant No. DIIU-DGA and the Programa 
Ibercaja-CAI  de Estancias de Investigaci\'on, Grant No. CB 5/21. J.~A. thanks the warm hospitality of the Universit\`a degli 
Studi di Padova and Istituto Nazionale di Fisica Nucleare. 


\bibliographystyle{hunsrt}
\bibliography{bibliography_alf}


\end{document}